\documentclass{article}
\usepackage{amsfonts}

\input{tcilatex}
\begin{document}

\title{Thermodynamic Basis for Odd Matter}
\author{\textbf{Martin Ostoja-Starzewski} \\
Department of Mechanical Science \& Engineering, \\
University of Illinois at Urbana-Champaign, USA}
\maketitle

\begin{abstract}
Continuum-type constitutive relations of odd matter need to be formulated
according to the second law of thermodynamics. Based on the primitive
thermodynamics of Edelen, a procedure admitting most general relations, is
outlined for heat-conducting fluids and solids. For viscous responses of odd
matter, the theory accounts for the irreversible and non-dissipative forces,
besides the hyperdissipative ones. For quasi-static responses, the theory
grasps the elastic and non-conservative forces besides those derivable from
the free energy, this being the realm of Cauchy elasticity beyond
hyperelasticity. In Cosserat-type odd matter, primitive thermodynamics also
accounts for curvature-torsion and couple-stress besides, respectively,
deformation and force-stress tensors. In both, classical and micropolar
cases, the theory grasps all possible couplings between the thermodynamic
velocity and force vectors, along with a full range of anisotropies of
hyperdissipative and hyperelastic responses (both linear and nonlinear).
\end{abstract}

\noindent \textit{Introduction}---One of the fundamental challenges in
continuum models of odd matter is the formulation of continuum-type
constitutive relations. Two phenomena characterize odd matter: its inability
to dissipate energy, and its inability to store energy [1-4]. In the first
case, we may speak of non-dissipative stresses, while in the second of
non-conservative stresses. Several models of such responses have been
introduced in the literature, see references in [4]. However, as pointed out
in [5], some of these models rely on constitutive relations improvised 
\textit{ad hoc} and are, in fact, expressed in field variables lacking
frame-invariance. In this paper, aiming at a formulation free of these
drawbacks, we work from the standpoint of \textit{primitive thermodynamics}
[6,7] to provide a minimal restriction on constitutive relations in
thermomechanics of fluid and solid odd matter.

As the starting point, odd matter equations should satisfy, just like in any
continuum model, the conservation laws for mass, linear momentum, angular
momentum, and energy as well as the second law of thermodynamics. In local
(in space and time) spatial (Eulerian) and material (Lagrangian)\
descriptions, this law reads in the Clausius-Planck form\ 
\begin{equation}
\rho \dot{s}+\mathbf{\nabla \cdot }\left( \frac{\mathbf{q}}{T}\right) \geq 
\frac{r}{T},\text{ \ \ }\rho _{0}\dot{s}_{0}+\mathbf{\nabla }\cdot \left( 
\frac{\mathbf{q}_{0}}{T}\right) \geq \frac{r_{0}}{T},
\end{equation}%
where $\rho $ is the mass density, $s$ the specific entropy (per unit
volume), $\mathbf{q}$\ the heat flux, $T$ the absolute temperature, and $r$
the heat source; subscript $_{0}$ refers to the reference configuration. The
overdot indicates the material time derivative $D/Dt$.

Additionally, all the conservation laws and constitutive relations should
not depend on the choice of frame used to describe them and ought to be
expressed in terms of objective quantities. Also called the principle of
material frame indifference [8], this requirement dictates that one must
write the constitutive relations in terms of strains (such as the
symmetrized displacement gradient) or deformation rate (the symmetrized
velocity gradient) in contradistinction to what occurs in some literature on
odd matter.

Since we work in the local description, each continuum point is endowed with
a local accompanying thermodynamic state. Thus, recalling the standard
relation for the free energy $\psi =u-Ts$ in terms of the internal energy $u$%
, (1) leads to the Clausius-Duhem inequality in spatial and material
descriptions%
\begin{equation}
\begin{array}{c}
-\rho \dot{\psi}-\rho s\dot{T}+\mathbf{\sigma }:\mathbf{d}-\frac{1}{T}%
\mathbf{\nabla }T\cdot \mathbf{q}\geq 0, \\ 
-\rho _{0}\dot{\psi}_{0}-\rho _{0}s_{0}\dot{T}+\mathbf{S}:\mathbf{\dot{E}}-%
\frac{1}{T}\mathbf{\nabla }T\cdot \mathbf{q}_{0}\geq 0.%
\end{array}%
\end{equation}%
The quantities in (2)$_{1}$ are the Cauchy stress tensor $\sigma _{ij}$, and
the deformation rate (i.e., symmetrized particle velocity gradient) $%
d_{ij}=v_{(i},_{j)}$, whereas in (2)$_{2}$ we see the the second
Piola-Kirchhoff stress tensor $\mathbf{S}$\ and the Green-St. Venant strain
tensor $\mathbf{E}$. All these quantities are fields parametrized by
location and time $\left( \mathbf{x},t\right) $ in the material domain of a
Euclidean space and time.

As is well known, the inequalities (2) may be written compactly in a
canonical form as%
\begin{equation}
\mathbf{V}\cdot \mathbf{Y}\left( \mathbf{V}\right) \geq 0,
\end{equation}%
where $\mathbf{V}$ and $\mathbf{Y}$ are vectors of \textit{thermodynamic
velocities} and \textit{thermodynamic forces}, respecitvely. The key problem
concerns the finding of a most general solution of (3), i.e. a constitutive
relation linking $\mathbf{Y}$ with $\mathbf{V}$. In this letter we show that
the primitive thermodynamics of Edelen [6,7] provides solutions covering the
inability of odd matter to dissipate power and/or store energy.

First, we treat heat conduction and, then, odd solid and fluid media within
the setting of classical continua, i.e. endowed with a translational degree
of freedom $\mathbf{u}$ ($=u_{i}\mathbf{e}_{i}$)\ per continuum particle; $%
\mathbf{e}_{i}$\ ($i=1,2,3$) form the orthonormal set of unit basis vectors
of the Euclidean space $\mathbb{E}^{3}$. Next, we extend the formulation to
micropolar solids and fluids, i.e. those endowed also with a rotational
degree of freedom $\mathbf{\varphi }$ ($=\varphi _{i}\mathbf{e}_{i}$). Both
index and symbolic tensor notations are used. The three-dimensional
formulation given here can always be reduced to planar situations.

\textit{Clausius-Duhem inequality in primitive thermodynamics}---Relying on
a theorem by Poincar\'{e} that any closed differential form on a star-shaped
domain is an exact differential, Edelen [6,7] showed that the most general
solution to the disspation inequality (3) is given by a so-called
decomposition theorem. According to it, every vector field $\mathbf{Y}$
which is of class $C^{1}$ in $\mathbf{V}$ and class $C^{2}$ in $\mathbf{w}$
admits the unique decomposition%
\begin{equation}
\mathbf{Y}=\nabla _{\mathbf{V}}\phi \left( \mathbf{V};\mathbf{w}\right) +%
\mathbf{U}\left( \mathbf{V};\mathbf{w}\right)
\end{equation}%
where $\phi \left( \mathbf{Y};\mathbf{w}\right) $ is a scalar-valued
function and the non-dissipative (or powerless) vector $\mathbf{U}\left( 
\mathbf{Y};\mathbf{w}\right) $ is such that%
\begin{equation}
\mathbf{V\cdot U}\left( \mathbf{V};\mathbf{w}\right) =0,\text{ \ \ \ }%
\mathbf{U\left( \mathbf{0};\mathbf{w}\right) =0.}
\end{equation}%
All odd, as opposed to even, constitutive responses are seen to be
represented by $\mathbf{U}\left( \mathbf{Y};\mathbf{w}\right) $. The role of 
$\mathbf{w}$\ will be displayed in examples below.

Effectively, the problem of solving the inequality (3) is reduced to the
problem of finding a scalar function $\phi \left( \mathbf{V};\mathbf{w}%
\right) $ such that%
\begin{equation}
\mathbf{Y}\cdot \nabla _{\mathbf{V}}\phi \left( \mathbf{V};\mathbf{w}\right)
\geq 0.
\end{equation}%
Here $\phi $ plays the role of a potential and is given by%
\begin{equation}
\phi =\int_{0}^{1}\mathbf{V}\cdot \mathbf{Y}(\tau \mathbf{V})d\tau ,
\end{equation}%
with $\mathbf{U}$ uniquely determined, for a given $\mathbf{Y}$, by%
\begin{equation}
U_{i}=\int_{0}^{1}\tau V_{j}\left[ \frac{\partial Y_{i}(\tau \mathbf{V})}{%
\partial V_{j}}-\frac{\partial Y_{j}(\tau \mathbf{V})}{\partial V_{i}}\right]
d\tau .
\end{equation}%
Also, the symmetry relations 
\[
\frac{\partial \left[ Y_{i}(\tau \mathbf{V})-U_{i}\right] }{\partial V_{j}}=%
\frac{\partial \left[ Y_{j}(\tau \mathbf{V})-U_{j}\right] }{\partial V_{i}} 
\]%
must hold, and these reduce to the classical Onsager reciprocity conditions $%
\partial Y_{i}(\tau \mathbf{V})/\partial V_{j}=$ $\partial Y_{j}(\tau
\partial \mathbf{V})/\partial V_{i}$\ if and only if $\mathbf{U}=\mathbf{0}.$

The part $\nabla _{\mathbf{V}}\phi \left( \mathbf{V};\mathbf{w}\right) $ of
the force vector in (9) may be termed \textit{hyperdissipative} after
Goddard [9,10], thus reflecting the fact that it is deriveable from a
dissipation potential $\phi $, just like the conservative part part of $%
\mathbf{Y}$ is deriveable from an elastic potential in a hyperelastic model.
Clearly, the non-dissipative force vector $\mathbf{U}\left( \mathbf{Y};%
\mathbf{w}\right) $ is needed for odd matter and, in the following, we
illustrate it on several examples. Goddard also obtained certain extensions
of Edelen's formulae through a generalization of the classical Gauss-Maxwell
construct.

\textit{Odd heat conduction}---Taking a rigid (undeformable) conductor,\
heat transfer is a purely dissipative process, where $\psi (\mathbf{e}%
)\equiv 0$\ and the Clausius-Duhem inequality reads 
\begin{equation}
-T^{-1}\mathbf{\nabla }T\cdot \mathbf{q}\geq 0.
\end{equation}%
This dictates $\mathbf{V}=\mathbf{\nabla }T$ and the thermodynamic force $%
\mathbf{Y}=-\mathbf{q}/T$, so that one seeks the most general solution $%
\mathbf{q}=\mathbf{q}\left( \mathbf{\nabla }T\right) $. Here, by a
mathematical analogy to strain in mechanics problems, we adopt the
temperature gradient\ as the argument of $\phi $, although a
complementary-type formulation with $\phi \left( \mathbf{q}\right) $ is also
possible. The simplest model is the Fourier heat conduction (with a thermal
conductivity tensor $\mathbf{\kappa }$, that needs to be positive
semi-definite)%
\begin{equation}
0\leq \phi \left( \mathbf{\nabla }T\right) =\mathbf{\kappa }:\mathbf{\nabla }%
T\cdot \mathbf{\nabla }T  \label{linear Fourier}
\end{equation}%
with a non-dissipative heat flux $\mathbf{U}$ ($=U_{i}\mathbf{e}_{i}$) given
as as%
\begin{equation}
U_{i}=D_{ij}T,_{j}\text{ \ with \ }D_{ij}=-D_{ji}.
\label{non-dissipative heat flux}
\end{equation}%
Clearly, taking an anti-symmetric tensor, $D_{ij}$, assures satisfaction of
(6)$_{2}$ to have an odd matter, while the dissipation potential $\phi $ may
be linear as in (10) or more general. In the latter case, one may proceed
by, say, taking the product $\mathbf{V}\cdot \mathbf{Y}(\mathbf{V})$
appearing\ in (7)\ to be a homogeneous function $D$ of degree $r$. One of
various other possibilities is the heat conduction with a relaxation time
leading to the hyperbolic heat transfer [5].

\textit{Odd solids with heat conduction}---Consider the case of a solid
storing the elastic energy, admitting small strains only $\mathbf{%
\varepsilon }:=\widehat{\mathbf{\nabla }}\mathbf{u}$ (symmetrized
displacement gradient $\varepsilon _{ij}=u_{(i},_{j)}$) and conducting heat.
Then, the Clausius-Duhem inequality (2)$_{1}$ is written as 
\begin{equation}
-\rho \dot{\psi}-\rho s\dot{T}+\mathbf{\sigma }:\mathbf{\dot{\varepsilon}}-%
\frac{1}{T}\mathbf{\nabla }T\cdot \mathbf{q}\geq 0,
\end{equation}%
where $\mathbf{\dot{\varepsilon}}:=\widehat{\mathbf{\nabla }}\mathbf{\dot{u}}
$\ is the small strain rate. The free energy is assumed in the form%
\begin{equation}
\psi =\psi \left( \mathbf{\varepsilon }\left( \mathbf{x},t\right) ,T\left( 
\mathbf{x},t\right) \right) ,
\end{equation}%
where $\varepsilon _{ij}$\ is the small strain tensor. By differentiating
(13), from (12) we find%
\begin{equation}
\left( \sigma _{ij}-\rho \frac{\partial \psi }{\partial \varepsilon _{ij}}%
\right) \dot{\varepsilon}_{ij}-\rho \left( s+\frac{\partial \psi }{\partial T%
}\right) \dot{T}-\frac{1}{T}T,_{i}\text{ }q_{i}\geq 0.
\end{equation}%
The vector space structure of this dissipation inequality involves:

(i) vector $\mathbf{V}$ of \textit{velocities} in a 10-dimensional vector
space%
\begin{equation}
\begin{array}{c}
\mathbf{V}=\left[ V_{k}\right] =\left[ \mathbf{\dot{\varepsilon}},\dot{T},%
\mathbf{\nabla }T\right] \\ 
\equiv \left[ \dot{\varepsilon}_{11},\dot{\varepsilon}_{22},\dot{\varepsilon}%
_{33},\dot{\varepsilon}_{23},\dot{\varepsilon}_{31},\dot{\varepsilon}_{12},%
\dot{T},T,_{1},T,_{2},T,_{3}\right] ,%
\end{array}%
\end{equation}

(ii) vector $\mathbf{Y}$ of \textit{forces} in a 10-dimensional vector space%
\begin{equation}
\mathbf{Y}=\left[ Y_{k}\right] =\left[ \mathbf{\sigma }-\rho \frac{\partial
\psi }{\partial \mathbf{\varepsilon }},-\rho s-\rho \frac{\partial \psi }{%
\partial T},\frac{-\mathbf{q}}{T}\right] ,
\end{equation}

(iii) vector $\mathbf{w}$ in a 7-dimensional space%
\begin{equation}
\mathbf{w}=\left[ w_{k}\right] =\left[ \mathbf{\varepsilon },T\right] \equiv %
\left[ \varepsilon _{11},\varepsilon _{22},\varepsilon _{33},\varepsilon
_{23},\varepsilon _{31},\varepsilon _{12},T\right] .
\end{equation}

It follows from (15)-(17) that\begingroup\renewcommand{\arraystretch}{1.9}%
\begin{equation}
\begin{array}{c}
\mathbf{\sigma }=\rho \displaystyle\frac{\partial \psi }{\partial \mathbf{%
\varepsilon }}+\frac{\partial \phi }{\partial \mathbf{\dot{\varepsilon}}}+%
\mathbf{U}^{\left( \mathbf{\sigma }\right) }, \\ 
\rho s=-\rho \displaystyle\frac{\partial \psi }{\partial T}+U^{\left(
s\right) }, \\ 
-\displaystyle\frac{\mathbf{q}}{T}=\displaystyle\frac{\partial \phi }{%
\partial \left( \mathbf{\nabla }T\right) }+\mathbf{U}^{\left( \mathbf{q}%
\right) }, \\ 
\mathbf{U}=\left( \mathbf{U}^{\left( \mathbf{\sigma }\right) },U^{\left(
s\right) },\mathbf{U}^{\left( \mathbf{q}\right) }\right) , \\ 
\mathbf{\dot{\varepsilon}}:\mathbf{U}^{\left( \mathbf{\sigma }\right) }+\dot{%
T}U^{\left( s\right) }+\mathbf{\nabla }T\cdot \mathbf{U}^{\left( \mathbf{q}%
\right) }=0.%
\end{array}%
\end{equation}%
\endgroup The vector $\mathbf{U}^{\left( \mathbf{q}\right) }$\ is given by (%
\ref{non-dissipative heat flux}).

If we now focus on the linear constitutive responses, these functions have
the~ forms\begingroup\renewcommand{\arraystretch}{1.8}%
\begin{equation}
\begin{array}{c}
\psi =\frac{1}{2}C_{ij}w_{i}w_{j},\text{ \ \ }i,j=1,2,...,7, \\ 
p=D_{ij}V_{i}V_{j},\text{ \ \ \ \ }D_{ij}V_{i}V_{j}\geq 0, \\ 
\phi =\int_{0}^{1}D_{ij}V_{i}V_{j}d\lambda ,\text{ \ \ }i,j=1,2,...,10.%
\end{array}%
\end{equation}%
\endgroup The positive semi-definiteness of the $\left[ D_{ij}\right] $\
matrix is assured by the Sylvester criterion.

Substituting (15)-(17) into (19) gives the expressions for $\psi $\ and $%
\phi $, accounting for all the possible individual and coupled field terms,%
\begingroup\renewcommand{\arraystretch}{2.0}%
\begin{equation}
\begin{array}{c}
\rho \psi =\frac{1}{2}C_{ijkl}^{\left( \mathbf{\varepsilon }\right)
}\varepsilon _{ij}\varepsilon _{kl}+\frac{1}{2}C^{\left( T\right)
}T^{2}+C_{ij}^{\left( \mathbf{\varepsilon }T\right) }\varepsilon _{ij}T\text{
\ \ \ \ } \\ 
\phi =\frac{1}{2}D_{ijkl}^{\left( \mathbf{\dot{\varepsilon}}\right) }\dot{%
\varepsilon}_{ij}\dot{\varepsilon}_{kl}+\frac{1}{2}D^{\left( \dot{T}\right) }%
\dot{T}^{2}+\frac{1}{2}D_{ij}^{\left( \dot{T}\right) }T,_{i}T,_{j} \\ 
+D_{ij}^{\left( \mathbf{\dot{\varepsilon}}\dot{T}\right) }\dot{\varepsilon}%
_{ij}\dot{T}+D_{i}^{\left( \dot{T}\mathbf{\nabla }T\right) }\dot{T}%
T,_{i}+D_{ijk}^{\left( \mathbf{\dot{\varepsilon}\nabla }T\right) }\dot{%
\varepsilon}_{ij}T,_{k}.%
\end{array}%
\end{equation}%
\endgroup While the above involves tensors of ranks 0 through 4, note the
following symmetries of tensors of ranks 2 through 4\begingroup%
\renewcommand{\arraystretch}{2.0}%
\begin{equation}
\begin{array}{c}
C_{ijkl}^{\left( \mathbf{\varepsilon }\right) }=C_{klij}^{\left( \mathbf{%
\varepsilon }\right) }=C_{jikl}^{\left( \mathbf{\varepsilon }\right)
}=C_{ijlk}^{\left( \mathbf{\varepsilon }\right) },\text{ \ \ }C_{ij}^{\left( 
\mathbf{\varepsilon }T\right) }=C_{ji}^{\left( \mathbf{\varepsilon }T\right)
}, \\ 
D_{ijkl}^{\left( \mathbf{\varepsilon }\right) }=D_{klij}^{\left( \mathbf{%
\varepsilon }\right) }=D_{jikl}^{\left( \mathbf{\varepsilon }\right)
}=D_{ijlk}^{\left( \mathbf{\varepsilon }\right) },\text{ \ \ }D_{ij}^{\left( 
\mathbf{\nabla }T\right) }=D_{ji}^{\left( \mathbf{\nabla }T\right) }, \\ 
D_{ij}^{\left( \mathbf{\dot{\varepsilon}}\dot{T}\right) }=D_{ji}^{\left( 
\mathbf{\dot{\varepsilon}}\dot{T}\right) },\text{ \ \ }D_{ijk}^{\left( 
\mathbf{\dot{\varepsilon}\nabla }T\right) }=D_{jik}^{\left( \mathbf{\dot{%
\varepsilon}\nabla }T\right) }.%
\end{array}%
\end{equation}%
\endgroup

To satisfy (18)$_{5}$, the admissible non-dissipative components of $\mathbf{%
U}$\ are set as

\begin{equation}
\begin{array}{c}
U_{ij}^{\left( \mathbf{\sigma }\right) }=B_{ijkl}^{\left( \mathbf{\sigma }%
\right) }\dot{\varepsilon}_{kl}\ \ \text{\ with \ \ }B_{ijkl}^{\left( 
\mathbf{\sigma }\right) }=-B_{klij}^{\left( \mathbf{\sigma }\right) }, \\ 
U^{\left( s\right) }=0, \\ 
U_{i}^{\left( \mathbf{q}\right) }=B_{ij}^{\left( \mathbf{q}\right) }T,_{j}%
\text{\ \ \ with \ \ }B_{ij}^{\left( \mathbf{q}\right) }=-B_{ji}^{\left( 
\mathbf{q}\right) }.%
\end{array}%
\end{equation}%
Note that the anti-symmetry (in the sense of major symmetry) of $%
B_{ijkl}^{\left( \mathbf{\sigma }\right) }$ as well as the anti-symmetry of $%
B_{ij}^{\left( \mathbf{q}\right) }$\ are crucial. Since we treat here the
case of a classical continuum (without microrotations), two minor symmetries
hold: $B_{ijkl}^{\left( \mathbf{\sigma }\right) }=B_{jikl}^{\left( \mathbf{%
\sigma }\right) }=B_{ijlk}^{\left( \mathbf{\sigma }\right) }$.

In view of (18), the complete constitutive responses are now determined
explicitly by combining the gradients of $\psi $\ and $\phi $\ with (22):%
\begin{equation}
\begin{array}{c}
\sigma _{ij}=C_{ijkl}^{\left( \mathbf{\varepsilon }\right) }\varepsilon
_{kl}+C_{ij}^{\left( \mathbf{\varepsilon }T\right) }T\text{ \ \ \ \ \ \ \ \
\ \ \ \ \ \ \ \ \ \ \ \ \ \ \ \ \ \ \ \ } \\ 
+D_{ijkl}^{\left( \mathbf{\dot{\varepsilon}}\right) }\dot{\varepsilon}%
_{kl}+D_{ij}^{\left( \mathbf{\dot{\varepsilon}}\dot{T}\right) }\dot{T}%
+D_{ijk}^{\left( \mathbf{\dot{\varepsilon}\nabla }T\right)
}T,_{k}+B_{ijkl}^{\left( \mathbf{\sigma }\right) }\dot{\varepsilon}_{kl} \\ 
\rho s=-C^{\left( T\right) }T+C_{ij}^{\left( \mathbf{\varepsilon }T\right)
}\varepsilon _{ij}\text{ \ \ \ \ \ \ \ \ \ \ \ \ \ \ \ \ \ \ \ \ \ \ \ \ \ }
\\ 
-\frac{1}{T}q,_{i}=D_{ij}^{\left( \dot{T}\right) }T,_{j}+D_{i}^{\left( \dot{T%
}\mathbf{\nabla }T\right) }\dot{T}+D_{ijk}^{\left( \mathbf{\dot{\varepsilon}%
\nabla }T\right) }\dot{\varepsilon}_{ij}+B_{ij}^{\left( \mathbf{q}\right)
}T,_{j}.%
\end{array}%
\end{equation}%
Note that constitutive laws of odd thermoelasticity or, simply, odd
elasticity and odd heat conduction alone are contained in (23) [5].

While the constitutive relations admit non-dissipative stresses, they do not
yet display non-conservative stresses, i.e. elastic but unable to store the
elastic energy. One could proceed \textit{ad hoc}, but, to produce such
stresses via primitive thermodynamics, we introduce an additive
decomposition of the stress tensor%
\begin{equation}
\mathbf{\sigma }=\mathbf{\sigma }^{\left( q\right) }+\mathbf{\sigma }%
^{\left( d\right) },
\end{equation}%
where $\mathbf{\sigma }^{\left( q\right) }$\ and $\mathbf{\sigma }^{\left(
d\right) }$\ are, respectively, the so-called \textit{quasi-conservative}
and \textit{dissipative} stresses. The most general solution for $\mathbf{%
\sigma }^{\left( d\right) }$ in a physically linear response may be derived
just like in the previous section. On the other hand, we have 
\begin{equation}
\sigma _{ij}^{\left( q\right) }=\rho \frac{\partial \psi }{\partial
\varepsilon _{ij}}+A_{ijkl}^{\left( \mathbf{\sigma }\right) }\varepsilon
_{kl}\ \ \text{\ with \ \ }A_{ijkl}^{\left( \mathbf{\sigma }\right)
}=-A_{klij}^{\left( \mathbf{\sigma }\right) }.
\end{equation}%
Clearly, $A_{ijkl}^{\left( \mathbf{\sigma }\right) }\varepsilon
_{kl}\varepsilon _{ij}=0$, showing that $A_{ijkl}^{\left( \mathbf{\sigma }%
\right) }\varepsilon _{kl}$\ is the non-conservative elastic stress.

A generalization to a geometrically and physically nonlinear case would
follow, with reference to (2)$_{2}$, by replacing $\mathbf{\sigma }$\ with $%
\mathbf{S}$\ and $\mathbf{\varepsilon }$\ with $\mathbf{E}$, while noting
the restrictions of writing linear constitutive relations in finite
elasticity.

\textit{Odd fluids with heat conduction}--- First, considering the case of a
viscous fluid conducting heat. Since the fluid does not store\ the free
energy ($\psi \equiv 0$), the internal energy density simplifies to $u=Ts$
and (2)$_{1}$ is written as%
\begin{equation}
-\rho s\dot{T}+\mathbf{\sigma }:\mathbf{d}-\frac{1}{T}\mathbf{\nabla }T\cdot 
\mathbf{q}\geq 0.
\end{equation}
This inequality can be cast in a canonical form $\mathbf{V}\cdot \mathbf{Y}%
\left( \mathbf{V}\right) \geq 0$\ with the identifications:

(i) vector $\mathbf{V}$ of velocities in a 10-dimensional vector space%
\begin{equation}
\mathbf{V}=\left[ X_{k}\right] =\left[ \mathbf{d},\dot{T},\mathbf{\nabla }T%
\right] ,
\end{equation}

(ii) vector $\mathbf{Y}$ of forces in a 10-dimensional vector space%
\begin{equation}
\mathbf{Y}=\left[ J_{k}\right] =\left[ \mathbf{\sigma },-\rho s,\frac{-q_{i}%
}{T}\right] ,
\end{equation}

(iii) vector $\mathbf{w\equiv 0}$.

Following the scheme of previous section, we have\begingroup%
\renewcommand{\arraystretch}{1.8}%
\begin{equation}
\begin{array}{c}
\mathbf{\sigma }=\rho \displaystyle\frac{\partial \phi }{\partial \mathbf{d}}%
+\mathbf{U}^{\left( \mathbf{\sigma }\right) }, \\ 
\rho s=-\rho U^{\left( s\right) }, \\ 
-\frac{1}{T}\mathbf{q}=\displaystyle\mathbf{U}^{\left( \mathbf{q}\right) },
\\ 
\mathbf{U}=\left( \mathbf{U}^{\left( \mathbf{\sigma }\right) },U^{\left(
s\right) },\mathbf{U}^{\left( \mathbf{q}\right) }\right) , \\ 
\mathbf{d}:\mathbf{U}^{\left( \mathbf{\sigma }\right) }+\dot{T}U^{\left(
s\right) }+\mathbf{\nabla }T\cdot \mathbf{U}^{\left( \mathbf{q}\right) }=0.%
\end{array}%
\end{equation}%
\endgroup Next, focusing on linear constitutive responses, on account of
(20)-(21), we need\begingroup\renewcommand{\arraystretch}{2.0}%
\begin{equation}
\begin{array}{c}
\phi =\frac{1}{2}D_{ijkl}^{\left( \mathbf{d}\right) }d_{ij}d_{kl}+\frac{1}{2}%
D^{\left( \dot{T}\right) }\dot{T}^{2}+D_{ij}^{\left( \mathbf{d\nabla }%
T\right) }d_{ij}T \\ 
+D_{ij}^{\left( \mathbf{d}\dot{T}\right) }d_{ij}\dot{T}+\frac{1}{2}%
D_{i}^{\left( \dot{T}\mathbf{\nabla }T\right) }\dot{T}T,_{i}+D_{ijk}^{\left( 
\mathbf{\dot{\varepsilon}\nabla }T\right) }d_{ij}T,_{k}.%
\end{array}%
\end{equation}%
\endgroup Again, note these symmetries of tensors of ranks 2 through 4%
\begingroup\renewcommand{\arraystretch}{2.0}%
\begin{equation}
\begin{array}{c}
D_{ij}^{\left( \mathbf{\dot{\varepsilon}}\dot{T}\right) }=D_{ji}^{\left( 
\mathbf{\dot{\varepsilon}}\dot{T}\right) },\text{ \ \ }D_{ijk}^{\left( 
\mathbf{\dot{\varepsilon}\nabla }T\right) }=D_{jik}^{\left( \mathbf{\dot{%
\varepsilon}\nabla }T\right) }, \\ 
D_{ijkl}^{\left( \mathbf{d}\right) }=D_{klij}^{\left( \mathbf{d}\right)
}=D_{jikl}^{\left( \mathbf{d}\right) }=D_{ijlk}^{\left( \mathbf{d}\right) },%
\text{ \ \ }D_{ij}^{\left( \mathbf{\nabla }T\right) }=D_{ji}^{\left( \mathbf{%
\nabla }T\right) }.%
\end{array}%
\end{equation}%
\endgroup To satisfy (29)$_{3}$, the admissible non-dissipative components
of $\mathbf{U}$\ are set just as in (22), providing $\dot{\varepsilon}_{kl}$%
\ is replaced by $d_{kl}$.

\textit{Odd micropolar solids with heat conduction}---When the symmetry of
the force-stress tensor $\mathbf{\sigma }$\ is broken, such as in active
matter, the couple-stress tensor $\mathbf{\mu }$ needs to be introduced to
ensure the satisfaction of angular momentum balance. To underline the lack
of symmetry, it is preferable to write $\mathbf{\tau }$\ instead of $\mathbf{%
\sigma }$ for the force-stress. The pair of stresses $\left( \mathbf{\tau },%
\mathbf{\mu }\right) $ is conjugate to the kinematic pair $\left( \mathbf{%
\dot{\gamma},\dot{\kappa}}\right) $, where $\mathbf{\dot{\gamma}}$ ($\gamma
_{ji}=u_{i},_{j}-\epsilon _{kji}\varphi _{k}$)\ is the strain rate and $%
\mathbf{\dot{\kappa}}$ ($\dot{\kappa}_{ji}=\dot{\varphi}_{i},_{j}$)\ the
curvature-torsion rate tensor. In the case of small $\mathbf{\gamma }$ and $%
\mathbf{\kappa }$, both Clausius-Duhem inequalities (2) become

\begin{equation}
-\rho \dot{\psi}-\rho s\dot{T}+\mathbf{\tau }:\mathbf{\dot{\gamma}}+\mathbf{%
\mu }:\mathbf{\dot{\kappa}}-\frac{1}{T}\mathbf{\nabla }T\cdot \mathbf{q}\geq
0.
\end{equation}

The vector space structure of this inequality is now made of a
19-dimensional $\mathbf{V}$ vector, a 19-dimensional $\mathbf{Y}$ vector,
and a 16-dimensional $\mathbf{w}$ vector. The derivation follows the same
steps as in odd solids without micropolar effects. The key thing to note is
that there are now two non-conservative tensors, a force-stress and a
couple-stress,%
\begin{equation}
\begin{array}{c}
U_{ij}^{\left( \mathbf{\tau }\right) }=B_{ijkl}^{\left( \mathbf{\tau }%
\right) }\dot{\gamma}_{kl}\text{\ \ \ with \ \ }B_{ijkl}^{\left( \mathbf{%
\tau }\right) }=-B_{klij}^{\left( \mathbf{\tau }\right) }, \\ 
U_{ij}^{\left( \mathbf{\mu }\right) }=B_{ijkl}^{\left( \mathbf{\mu }\right) }%
\dot{\kappa}_{kl}\text{\ \ \ with \ \ }B_{ijkl}^{\left( \mathbf{\mu }\right)
}=-B_{klij}^{\left( \mathbf{\mu }\right) }.%
\end{array}%
\end{equation}%
Since $B_{ijkl}^{\left( \mathbf{\tau }\right) }$ and $B_{ijkl}^{\left( 
\mathbf{\mu }\right) }$\ are anti-symmetric in the ($ij$) and ($kl$) pairs, $%
U_{ij}^{\left( \mathbf{\tau }\right) }\dot{\gamma}_{ij}=0$ and $%
U_{ij}^{\left( \mathbf{\mu }\right) }\dot{\kappa}_{ij}=0$\ hold irrespective
of the fact that $\dot{\gamma}_{ij}$ and $\dot{\kappa}_{ji}$\ generally lack
symmetry in $i$ and $j$. This picture may be adapted to elastic
non-conservative phenomena like, say, visco/elasto-plastic.

In odd micropolar viscous fluids, the starting point is given by the
Clausius-Duhem inequality (2)$_{1}$ generalized to 
\begin{equation}
-\rho \dot{\psi}-\rho s\dot{T}+\tau _{kl}\left( v_{l},_{k}-\epsilon _{klr}%
\dot{\varphi}_{r}\right) +\mu _{kl}\dot{\varphi}_{l},_{k}-\frac{1}{T}%
T,_{i}q_{i}\geq 0.
\end{equation}%
The constitutive relations are expressed through the hyperdissipative force-
and couple-stresses derived as gradients of the (positive semi-definite)
dissipation function, while the irreversible non-dissipative stresses have
forms analogous to (33).

\textit{Conclusion}---Edelen's representation provides the most general
solution of the Clausius-Duhem inequality that is written as a scalar
product of the vector of thermodynamic velocities with the vector of
thermodynamic forces. As a result, primitive thermodynamics accounts for
irreversible and non-dissipative forces, besides the hyperdissipative ones
in the odd matter. Concerning the quasi-static responses, besides
hyperelastic forces, the elastic and non-conservative forces are identified
in an analogous manner for odd solid-like matter $-$ this is the\ realm of
Cauchy elasticity beyond hyperelasticity where the stress state is a
function of strain state: $\mathbf{\sigma }=\mathbf{\sigma }\left( \mathbf{%
\varepsilon }\right) $. As explicit examples, heat-conducting odd solids and
odd fluids are worked out, with the odd heat conduction in a rigid heat
conductor obtained as a special case.

In a Cosserat-type continuum where (i) a continuum point's translation is
described by a rotation (in addition to a translation) and (ii) the stress
state is grasped by a couple-stress (in addition to a force-stress), the
Clausius-Duhem inequality is enlarged to also account for additional
micropolar effects. The same approach as in the classical (non-polar) case
is to be adopted to determine all possible couplings between the
thermodynamic velocity and force vectors. Also, a full range of different
anisotropies of hyperdissipative and hyperelastic responses (both linear and
nonlinear) can be grasped, whereas the non-dissipative and/or
non-conservative $\mathbf{U}$ vector relies on the anti-symmetry (in the
sense of major symmetry) of the linear response tensor. For odd solid
matter, this is the\ realm of Cauchy micropolar elasticity: $\mathbf{\sigma }%
=\mathbf{\sigma }\left( \mathbf{\varepsilon },\mathbf{\kappa }\right) $, $%
\mathbf{\mu }=\mathbf{\mu }\left( \mathbf{\varepsilon },\mathbf{\kappa }%
\right) $. A vast range of many other constitutive responses of odd matter
can be described using the thermodynamic basis outlined here.

\bigskip

\textit{Acknowledgment: }We benefitted from comments of J.D. Goddard of UC
San Diego.

\bigskip

[1] D. Banerjee, V. Vitelli, F. J\"{u}licher, and P. Sur\'{o}wka, Active
viscoelasticity of odd Materials, \textit{Phys. Rev. Lett.} \textbf{126},
138001 (2021). Erratum \textit{Phys. Rev. Lett.} \textbf{127}, 189901 (2021).

[2] R. Lier, J. Armas, S. Bo, C. Duclut, F. J\"{u}licher, and P. Sur\'{o}%
wka, Passive odd viscoelasticity, \textit{Phys. Rev. E} \textbf{105}, 054607
(2022).

[3] P. Sur\'{o}wka, A. Souslov, F. J\"{u}licher, and D. Banerjee, Odd
Cosserat elasticity in active materials, \textit{Phys. Rev. E} \textbf{108},
064609 (2023).

[4] M. Fruchart, C. Scheibner, and V. Vitelli, Odd viscosity and odd
elasticity, \textit{Annu. Rev. Condens. Matter Phys}. \textbf{14}, 471
(2023).

[5] M. Ostoja-Starzewski and P. Sur\'{o}wka, Generalizing odd elasticity
theory to odd thermoelasticity for planar materials, \textit{Phys. Rev. B} 
\textbf{109}, 064107 (2024).

[6] D.G.B. Edelen, On the existence of symmetry relations and dissipative
potentials. \textit{Arch. Rat. Mech. Anal}. \textbf{51}, 218-227 (1973).

[7] D.G.B. Edelen, Primitive thermodynamics:\ A new look at the
Clausius-Duhem inequality. \textit{Int. J. Eng. Sci.} \textbf{12}, 121-141
(1974).

[8] C. Truesdell, and W. Noll, \textit{The Non-Linear Field Theories of
Mechanics}. In: Fl\"{u}gge S (ed) \textit{Encyclopedia of Physics} \textbf{%
III/3}. Springer, Berlin (Springer, Berlin, 1965).

[9] J.D. Goddard, Edelen's dissipation potentials and the visco-plasticity
of particulate media. \textit{Acta Mech}. \textbf{225}, 2239-2259 (2014).

[10] J. Goddard, Continuum modeling of granular media. \textit{Appl. Mech.
Rev}. \textbf{66}, 050801 (2014).

\end{document}